\documentclass[a4paper,11pt]{article}
\usepackage{jinstpub} % for details on the use of the package, please
\usepackage{subfigure}

\usepackage{booktabs}

\usepackage{lineno}

\title{Design and Operation of the PandaX-4T High Speed Ultra-high Purity Xenon Recuperation System}

\author[a,b,c]{Zhou Wang,}
\author[a]{Wenbo Ma,}
\author[a,b,1]{Tao Zhang\note{Corresponding author.},}
\author[a,b]{Li Zhao,}
\author[c]{Shuaijie Li,}
\author[c]{Xiangyi Cui,}
\author[a,b,c]{Jianglai Liu,}
\author[d]{Changbo Fu,}
\author[e]{Yonglin Ju,}
\author[f,g]{Qing Lin,}
\author[c]{Xiaohua Chen,}
\author[e]{Xiuli Wang}

\affiliation[a]{ School of Physics and Astronomy, Shanghai Jiao Tong University, MOE key Laboratory for Particle Astrophysics and Cosmology,\\Shanghai 200240, China}
\affiliation[b]{Shanghai Jiao Tong University Sichuan Research Institute,\\Chengdu 610000, China}
\affiliation[c]{Tsung-Dao Lee Institute, Shanghai Jiao Tong University, \\Shanghai 200240, China}
\affiliation[d]{ Institute of Modern Physics, Fudan University, Key Lab of Nuclear Physics and Ion-beam Application (MOE), \\Shanghai 200240, China}
\affiliation[e]{Institute of Refrigeration and Cryogenics, Shanghai Jiao Tong University\\Shanghai 200240, China}
\affiliation[f]{State Key Laboratory of Particle Detection and Electronics, University of Science and Technology of China \\Hefei 230026, China}
\affiliation[g]{Department of Modern Physics, University of Science and Technology of China \\Hefei 230026, China}

\emailAdd{tzhang@sjtu.edu.cn}

\abstract{In order to recuperate the ultra-high purity xenon from PandaX-4T dark matter detector to high-pressure gas cylinders in emergency or at the end-of-run situation, a high speed ultra-high purity xenon recuperation system is designed and developed. This system includes a diaphragm pump, the heat management system, the main recuperation pipeline, the reflux pipeline, the auxiliary recuperation pipeline and the automatic control system. The liquid xenon in the detector is vaporized by the heat management system, and the gaseous xenon is compressed to 6 MPa at the flow rate of 200 standard litres per minute (SLPM) using the diaphragm compressor. The high-pressure xenon is filled into 128 gas cylinders via the main recuperation pipeline. During the recuperation, the low pressure and temperature conditions of 2 $\sim$ 3 atmospheres and 178 $\sim$ 186.5 K in PandaX-4T dark matter detector are kept by the cooperation of the main recuperation pipeline, reflux pipeline and the auxiliary recuperation pipeline to guarantee the safety, and the purity of the recuperated xenon gas is measured to ensure no contamination happened. The development of the high speed ultra-high purity xenon recuperation system is important for the operation of large-scale dark matter detectors with the requirements of strict temperature and pressure environment and low background.
}


\begin{document}
\maketitle
\flushbottom

\section{Introduction}
\label{sec:intro}

Dark matter is a kind of mysterious, nonluminous newly unknown matter not involved in electromagnetic and strong interactions, the mass of dark matter is five times more than normal matter. In recent years, the detection of dark matter is one of the most advanced research topics in particle physics~\cite{1}.

Xenon (Xe) is regarded as one of the most attractive medium for dark matter detectors because of the advantages of low energy threshold, distinguishable energy resolution, high ionization yield and high light yield. Furthermore, there is no long-lived radioactive isotope so the background of xenon itself is very low~\cite{2,3,4}. The PandaX project consists of a series of xenon-based experiments located at the China Jinping Underground Laboratory (CJPL)~\cite{5,6,7}. The PandaX-I and PandaX-II are the two completed experiments with targets of 120 kg and 580 kg, respectively. PandaX-4T is the new generation experiment, which has a target of 4-ton liquid xenon~\cite{8}. 

PandaX dark matter detector works at the vapour-liquid saturation point (178K versus $ 2\times10^{5}$ Pa, the absolute pressure is used in this paper). It is a gas-liquid dual-phase xenon Time Projection Chamber (TPC)~\cite{9}. The operation duration of the detector lasts for years after the completion of the commissioning run to increase the exposure of the detection data, thus to reduce the detection error and improve the reliability and credibility of the results~\cite{6}. During the long-term dark matter detection process, if the emergency conditions appear such as refrigeration equipment failure or the outer vacuum increase, the liquid xenon in the detector would vaporize and make the pressure increase because of thermal leakage, leading to the damage of the photo-multiplier tubes (PMT, HAMAMATSU), which are the critical detection devices in the dark matter detector and their maximum bearing pressure is $ 3.5\times10^{5}$ Pa. Furthermore, the xenon would leak from the dark matter detector because of over-pressure, thus causes a serious accident. In the emergency situation, the vaporized xenon gas needs to be recuperated quickly to avoid the accident. On the other hand, a high speed xenon recuperation system is needed to recuperate $\sim$6 tons of xenon from the detector to the high-pressure xenon cylinders for storage at the end of the experiment. In the meanwhile, the xenon in the detector would be frozen if the pressure inside is too low. As a result, It is better to control the pressure in the detector between $2\times10^{5}$ Pa and $3\times10^{5}$ Pa during the recuperation.

During the recuperation, it is necessary to ensure there is no leakage during the recuperation process because Xe is a kind of expensive noble gas. In addition, the contamination of Xe needs to be minimized to meet the requirement of low background of the detector for the next step of the dark matter experiment. 

In previous experiments, the Xe in the dark matter detector was recuperated into several aluminum bottles cooled by liquid nitrogen due to the pressure difference in PandaX-I, PandaX-II and Xenon 100 dark matter detection experiments~\cite{5,10,11}. The flow rate of this recuperation method is 20 $\sim$ 30 SLPM, and it would take about two months to recuperate $\sim$6 tons of Xe from PandaX-4T detector. Xenon 1T dark matter detection experiment utilized a spherical liquid xenon storage tank with a diameter of 2.1 m called ReStoX to recuperate liquid xenon from the detector directly, which could reach the flow rate to 1$\sim$2 tons per hour. In ReStoX, liquid nitrogen lines are welded to the outer surface of the inner vessel to cool down the sphere, with thin stainless-steel fins inside the volume additionally to increase the heat exchange, and 7.6 tons of liquid xenon can be stored~\cite{12}. The diaphragm compressor was used in LUX-ZEPLIN (LZ) dark matter detection experiment to transfer xenon gas from the detector to xenon gas bottles at the flow rate of 200 SLPM~\cite{13}.

PandaX-4T high speed ultra-high purity (UHP) xenon recuperation system utilizes diaphragm compressor to compress the xenon gas in the dark matter detector from  2 $\sim$ 3$ \times10^{5}$ Pa to 6 MPa, then fill it into 128 standard high-pressure gas cylinders of 40 L for each at the flow rate of 200 SLPM. This system has recuperated 5.6 tons of xenon from PandaX-4T detector successfully after its first scientific run, and the experimental data analysis is meaningful for the optimization of the recuperation system. The upgrade of the high speed Xe recuperation and heat management technique is important for the next generation of the detectors of dozens of tons scale, and even hundreds of tons scale. In this paper, the design and main parameters of the newly high speed UHP xenon recuperation system is introduced in section~\ref{sec:2}. The hardware setup and the operation process are described in section~\ref{sec:3}. The operation process of the system and results are discussed in section~\ref{sec:4}, before we conclude in section~\ref{sec:5}.

\section{Design and construction of the recuperation system}
\label{sec:2}

It is important to design the pipeline system and working process for the high speed UHP xenon recuperation system. In order to achieve the purpose of large recuperate flow rate of 200 SLPM, a diaphragm compressor is used to compress 2 $\sim$ 3$\times10^{5}$ Pa xenon in the dark matter detector to 6 MPa then push it into high-pressure bottle system at the temperature of 17 $^{\circ}$C~\cite{14}. In the recuperation process, it is required neither leakage nor contamination to ensure the quantity and purity of the recuperated xenon.

The design and construction targets of PandaX-4T UHP xenon recuperation system are listed below:

\begin{itemize}

\item The pressure of the xenon in the detector is kept between $ 2\times10^{5}$ Pa and $ 3\times10^{5}$ Pa in order to ensure the safety of the dark matter detector.

\item The recuperation flow rate reaches to 200 SLPM.

\item The pressure after the diaphragm compressor reaches 6 MPa.

\item The system operates stably and continuously.

\item There is neither leakage nor contamination during the recuperation process.
\end{itemize}

\subsection{Ultra-high purity xenon diaphragm compressor}
\label{sec:2:1}

\begin{figure}[htbp]
\centering 
\includegraphics[width=.7\textwidth]{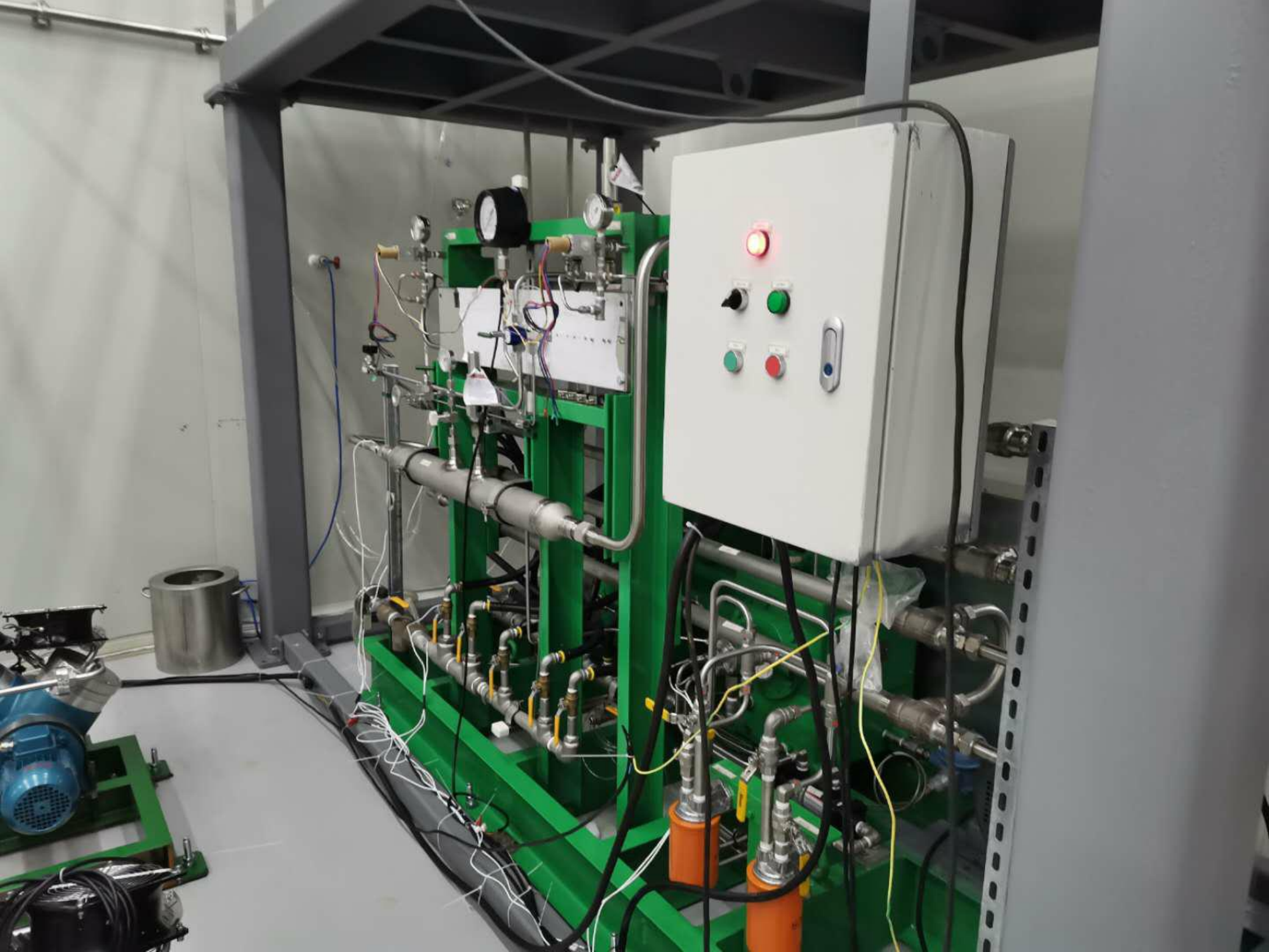}
\caption{\label{fig:2:1:1} The picture of the PPI UHP xenon diaphragm compressor.}
\end{figure}

The critical equipment of the recuperation system is the UHP xenon diaphragm compressor (Model 4LX-180-132, water cooling, PPI, Inc.), the external structure of the compressor is shown in figure ~\ref{fig:2:1:1}. The compressor is double stage compression with two membrane heads for the UHP xenon according to the customization. Metal membranes are adopted to ensure the cleanliness of the medium xenon gas. 

\begin{figure}[htbp]
\centering 
\includegraphics[width=.7\textwidth]{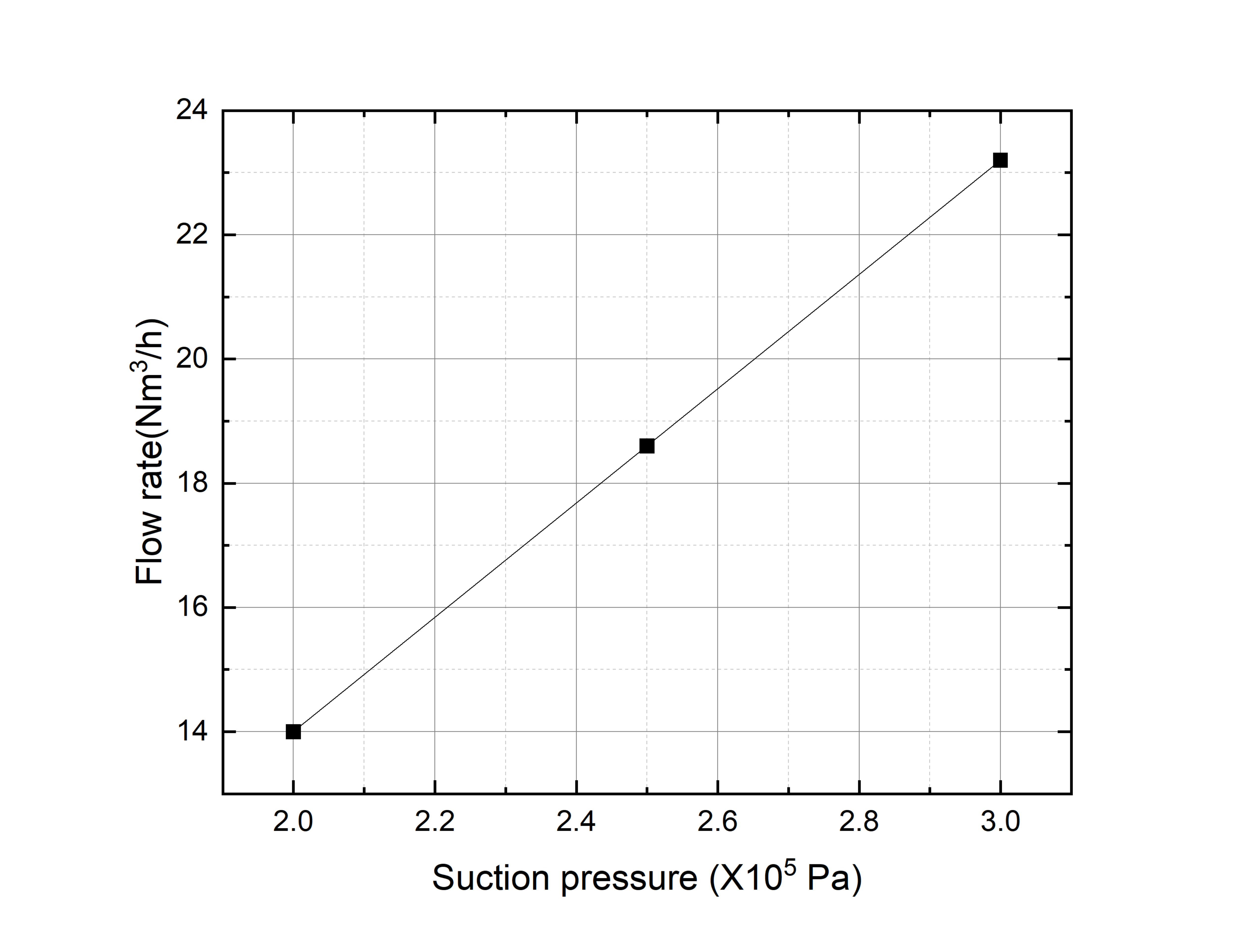}
\caption{\label{fig:2:1:2} The flow chart of the PPI UHP xenon diaphragm compressor.}
\end{figure}

The inlet pressure of the diaphragm compressor is between $ 2\times10^{5}$ Pa and $ 3\times10^{5}$ Pa, and the corresponding xenon gas flow is from $ 14 Nm^{3}/h$ to $ 23.2 Nm^{3}/h$ (233.33 SLPM to 386.67 SLPM), according to figure ~\ref{fig:2:1:2}. The compression ration of the first stage is 4.3, corresponding to the inter-stage pressure of 1.2 MPa. The compression ratio of the second stage is 4.7, and the outlet pressure of the diaphragm compressor could reach 6 MPa.

\subsection{Heat management system}
\label{sec:2:2}

The volume of PandaX-4T dark matter detector is 3 $m^{3}$, in which the volumes of gas and liquid xenon are 1 $ m^{3}$ and 2 $ m^{3}$ initially, with the corresponding mass of 18.68 kg (gaseous xenon density is 18.68 $ kg/m^{3}$ at 178 K and $ 2\times10^{5}$ Pa) and $\sim$6 t (liquid xenon density is 2853.9 $ kg/m^{3}$ at 178 K and $ 2\times10^{5}$ Pa), respectively. It would take 7.5 min to recuperate xenon gas in dark matter detector to under $ 1\times10^{5}$ Pa after starting the diaphragm compressor, and the negative pressure would cause the risk of freezing the liquid xenon in the detector, thus destroy PMTs. As a result, it is important to design and construct a heating system to vaporize the liquid xenon to avoid low pressure in the detector.

 Ethanol heating circulation system is chosen to be utilized in PandaX-4T recuperation system. There are two advantages to choose ethanol as the heating medium: firstly, the melting point of ethanol is 159 K at $ 1\times10^{5}$ Pa, so liquid circulation could be kept without freezing when exchanging heat with the liquid xenon in the detector of the low temperature of 178 K. Secondly, The ethanol heating is chosen as the heat management method instead of electrical heating because ethanol is clean and its radiation is low, which is important to the high-sensitive dark matter detector. The tightness of the ethanol heating circulation system is checked strictly to ensure the operation safety.

\begin{figure}[htbp]
\centering 
\includegraphics[width=.9\textwidth]{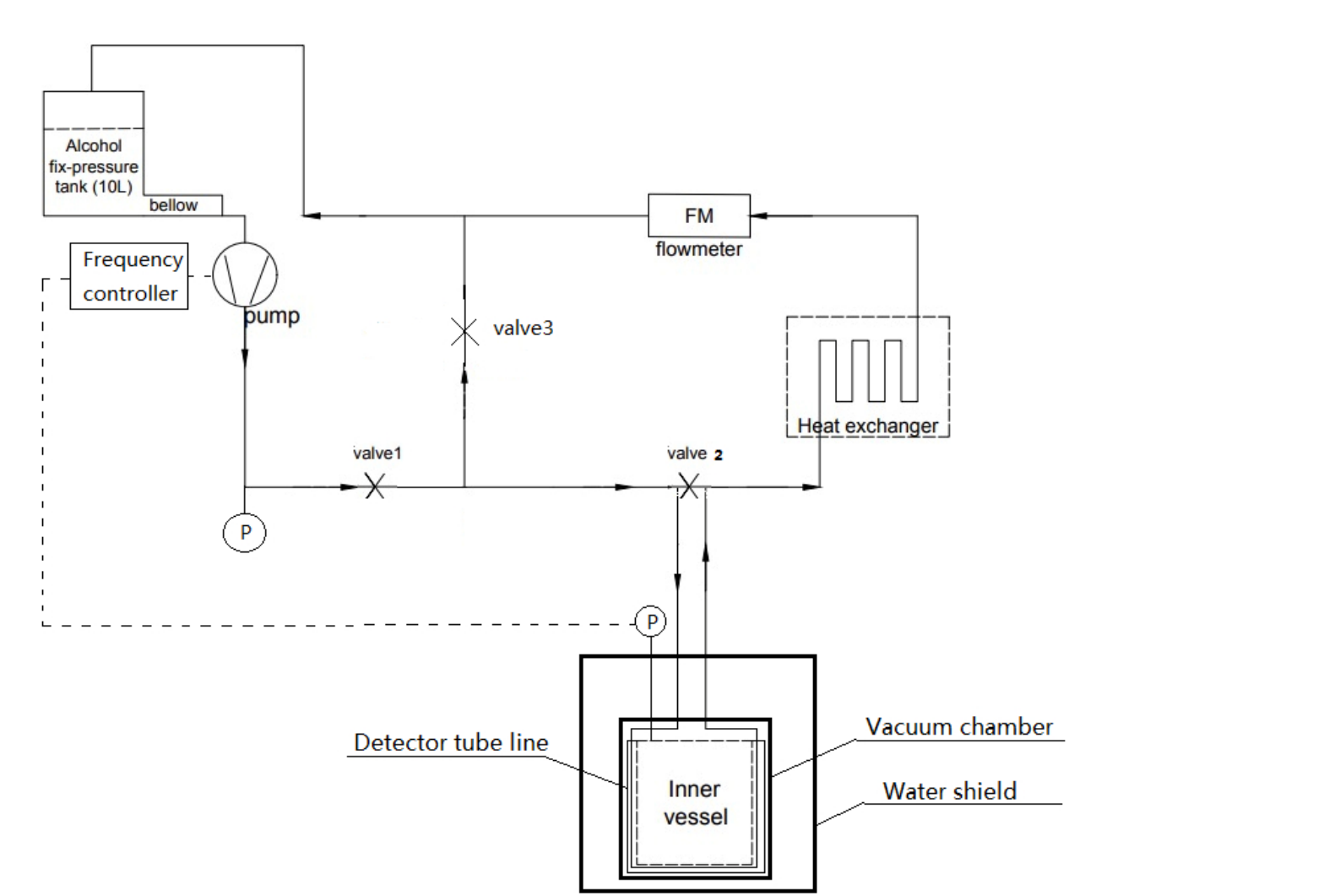}
\caption{\label{fig:2:2:1} The flow diagram of the ethanol circulation heating system of the UHP xenon recuperation system.}
\end{figure}

The flow diagram of the ethanol heating circulation system is shown in figure ~\ref{fig:2:2:1}. The external tube line in room temperature and the detector tube line in low temperature are included in this system. The external tube line consists of an ethanol storage tank, a variable-frequency gear circulation pump, an explosion-proof flow meter, a heat exchanger and several valves. The devices mentioned above are located at room temperature environment outside the detector. The detector tube line is the U-shaped rectangular pipeline with a cross section of 30 mm $\times$ 10 mm, welded on the outer wall of the inner detector vessel, which is in the outer vacuum chamber of the dark matter detector.

The ethanol in the ethanol storage tank is driven into the detector tube line by the circulation pump. The pressure at the outlet of the circulation pump is monitored by a pressure gauge. The ethanol at room temperature in the tube line exchange heat with the liquid xenon, via the stainless steel wall with 6 mm thickness of the inner chamber of the detector, and the liquid xenon is vaporized. The thermal power of the heat exchange Q at the vertical detector tube line is calculated as below:

\begin{equation}
\label{eq1:1}
Q =  \int_{0}^{A}Q_{x}dA \,
\end{equation}
\begin{equation}
\label{eq1:2}
Q_x = h_{x}\cdot(t_{x}-t_{w}) \,
\end{equation}
\begin{equation}
\label{eq1:3}
Re = \frac{ud}{\upsilon} \,
\end{equation}
\begin{equation}
\label{eq1:4}
h_x = 0.332\cdot\frac{\lambda}{x}Re^{1/2}Pr^{1/3} \,
\end{equation}
\begin{equation}
\label{eq1:5}
Pr = \frac{\mu c_{p}}{\lambda} \,
\end{equation}

where $Re$ is Reynolds number, and its value is 80 at max designed ethanol flow rate of 1 LPM, which is less than 2300, so the ethanol flow belongs to laminar flow.

In addition, $A$ indicates the contact area, $m^2$; $h_{x}$ is convective heat transfer coefficient at x position, $W/(m^2\cdot K)$; $t_{x}$ is the ethanol temperature at x position, K; $t_{w}$ is the wall temperature of 178 K; $\lambda$ is heat conductivity of the ethanol which is 0.178 $W/(m\cdot K)$; $Pr$ is Prandtl number; $u$ is the flow velocity of ethanol, m/s, d is the height of the rib tube which is $10^{-2}$ m; $\upsilon$ is the kinematic viscosity of the ethanol which is 7$\times10^{-6} m^2/s$, $\mu$ is the dynamic viscosity of the ethanol which is 1.15$\times10^{-2} Pa\cdot s$, and $c_p$ is specific heat at constant pressure of the ethanol which is 2 $kJ/(kg\cdot K)$.

Eqs.\eqref{eq1:1}-\eqref{eq1:5} are calculated by finite element analysis method. According to the calculation, even when the flow rate of the ethanol reaches to 3 LPM, the maximum heat flux transfer to the liquid xenon does not exceed 5 $W/cm^{2}$, which would not lead to local violent boiling, so the ethanol heating method is relatively mild and safe. The heating power required to vaporize the liquid xenon at the recuperation rate of 200 SLPM is 1.666 kW.

On the other hand, the effective heating area is maximum at the early stage of the recuperation due to the high liquid xenon level in the detector. The ethanol could exchange heat with the liquid xenon sufficiently, so the heating power is proportional to the ethanol flow rate. The heating power is about 1 kW corresponding to the ethanol flow rate of 0.3 SLPM at this stage according to the simulation calculation. The ethanol flow rate increases to enlarge the heat flux along the recuperation process because of the decreasing of the effective heat exchange area, caused by liquid xenon level dropping. At the end of the recuperation, the heating power needs to be maintained by increasing the ethanol circulation flow rate rapidly.

The low temperature ethanol from the detector tube line recovers to room temperature after flowing into the heat exchanger to exchange heat with the air. The tube line between detector tube line and external tube line is merged in the water shield which help warm the ethanol up, but there is no obvious freeze happened in the water because of its large heat capacity. There are two temperature sensors (PT100) set before and after the heat exchanger to monitor the temperature of the ethanol at the inlet and outlet of the heat exchanger. Then the room temperature ethanol is filled into the storage tank and injected to the detector tube line for vaporizing liquid xenon via the pump, and the circulation is completed.

The circulation flow range of the ethanol is from 0 to 1 LPM, which is adjusted by the variable-frequency gear circulation pump. A bypass tube line is used to control the flow rate of ethanol by adjusting the bypass valve as well. The frequency of the gear circulation pump is controlled according to the internal pressure of the detector. The specific control situation is illustrated in section~\ref{sec:3}.

The circulation system stops when the pressure in the detector is high, the residual ethanol left in the detector tube line is about 1.3 L, corresponding to the residual heat of less than $ 2.6\times10^{5}$ J, and it causes the maximum rising pressure of the detector of $ 0.45\times10^{5}$ Pa, which indicates the thermal inertia of the ethanol heating method is small and the system is easy to be controlled.

\begin{figure}[htbp]
\centering 
\includegraphics[width=.7\textwidth]{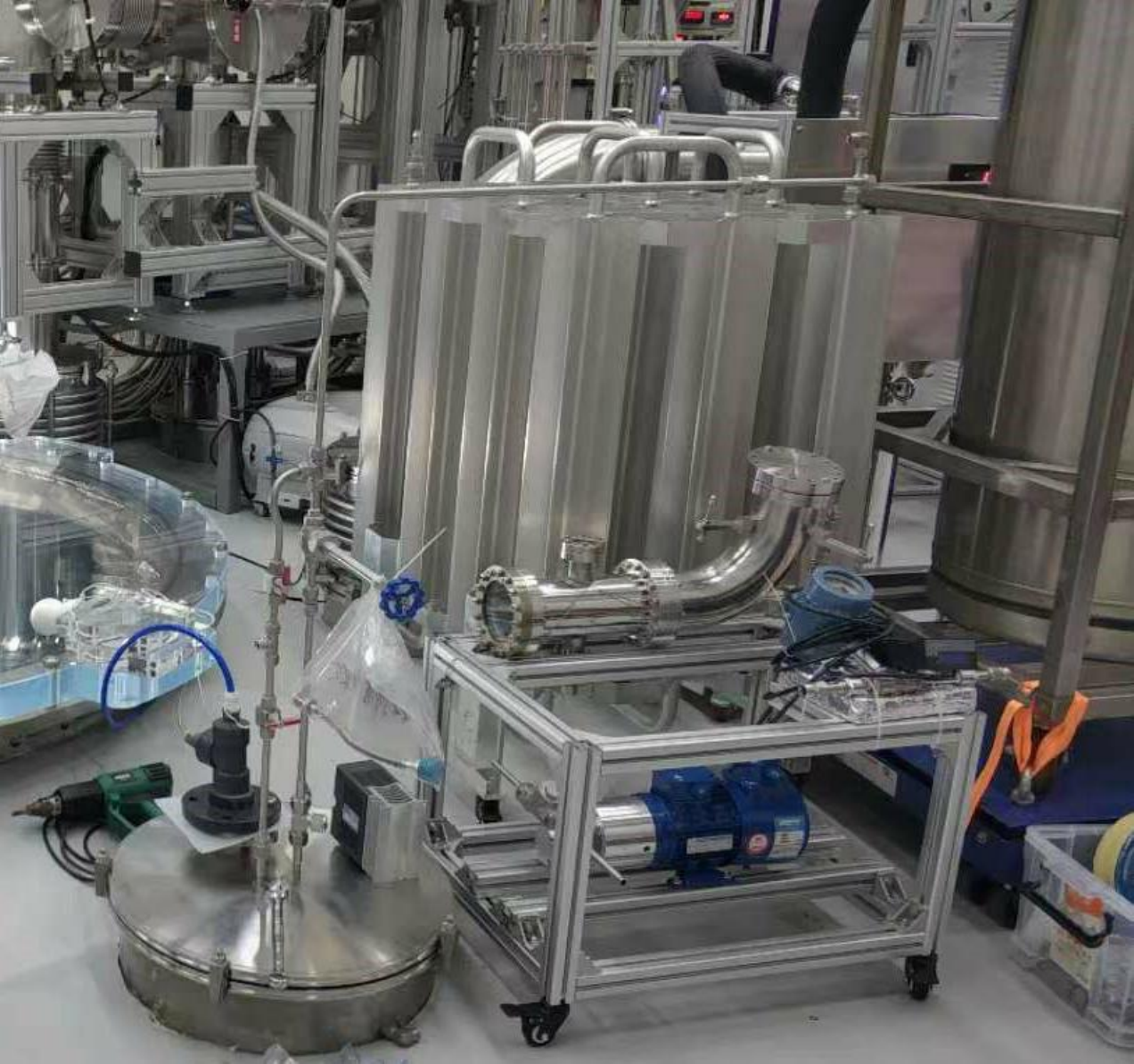}
\caption{\label{fig:2:2:2} The picture of the ethanol circulation heating system of the UHP xenon recuperation system.}
\end{figure}

The picture of the ethanol circulation heating system of the UHP xenon recuperation system is shown in figure ~\ref{fig:2:2:2}.

\subsection{Ultra-high purity xenon recuperation pipeline system}
\label{sec:2:3}

\begin{figure}[htbp]
\centering 
\includegraphics[width=1.\textwidth]{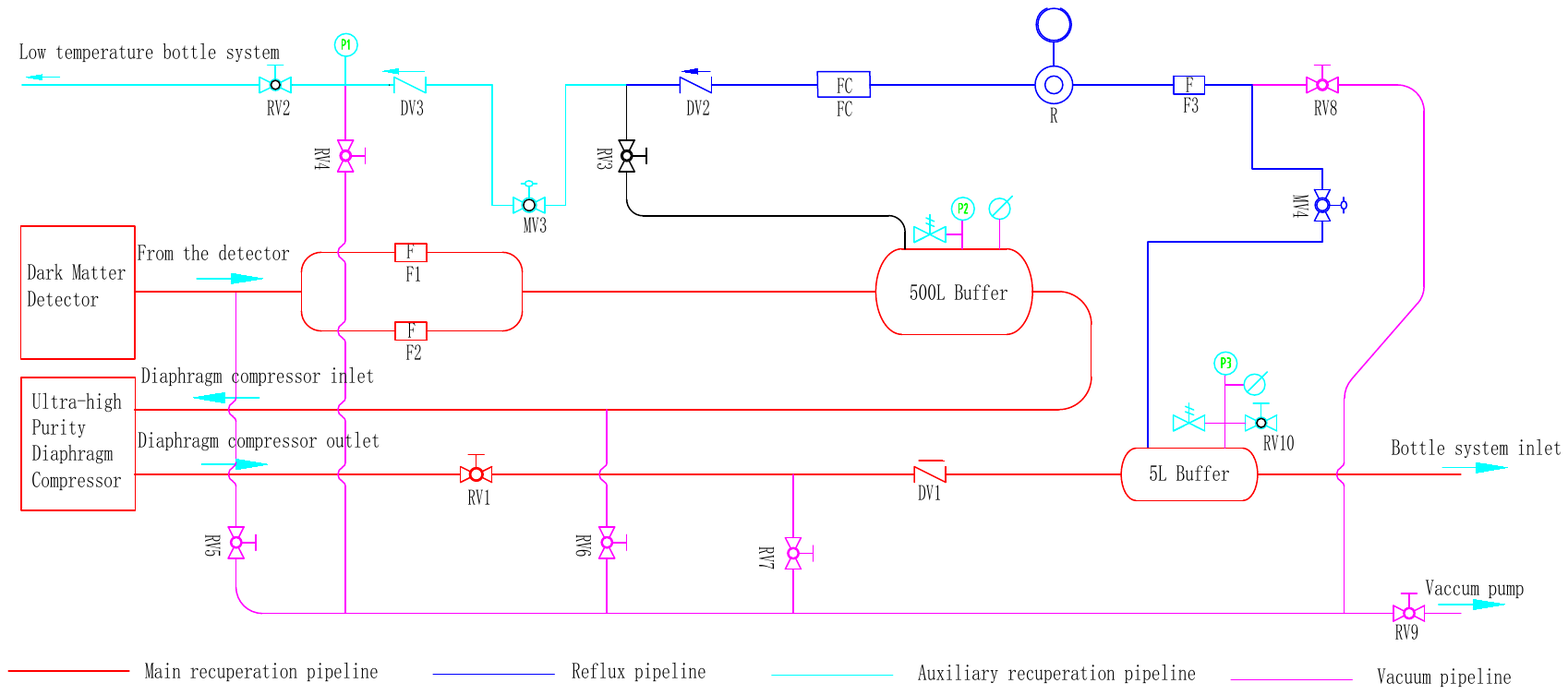}
RV-Bellow Valve; DV-Check Valve; F-Filter; FC-Flow Controller; R-Regulator; MV-Pneumatic Valve; P-Pressure Gauge
\caption{ \label{fig:2:3:1}The flow diagram of the UHP xenon recuperation system }
\end{figure}

The UHP xenon recuperation pipeline system includes the main xenon recuperation pipeline, reflux pipeline, auxiliary recuperation pipeline and vacuum pipeline, as shown in figure ~\ref{fig:2:3:1}.

In  figure ~\ref{fig:2:3:1}, the red line is the main xenon recuperation pipeline, whose purpose is to compress the xenon gas in the detector from 2 $\sim$ 3 $\times10^{5}$ Pa to 6 MPa at the rate of 200 SLPM via the diaphragm compressor, and fill it into the gas bottle system.

The cyan line indicates the auxiliary recuperation pipeline, which is designed to deal with the situation of relatively high pressure in the dark matter detector. The valve of the auxiliary recuperation pipeline opens when the pressure in the detector increases to $ 2.5\times10^{5}$ Pa, in order to assist recuperating the xenon gas from the detector at the flow rate of 50 SLPM.

The blue line illustrates the reflux pipeline, which is utilized to deal with the situation of relatively low pressure in the dark matter detector. The valve of the reflux pipeline opens to replenish xenon from the high pressure pipeline after the diaphragm compressor to the low pressure pipeline before it via a regulator, when the pressure in the detector decreases to $ 2\times10^{5}$ Pa.

The pink line indicates the vacuum pipeline. The whole recuperation system is required to be vacuum pumped and leak checked using helium (He) to ensure the tightness and cleanliness before starting. The leakage rate and the vacuum are required to be lower than $1\times10^{-11} Pa \cdot m^{3}/s$ and at $ 1\times10^{-4}$ Pa level, respectively.

\begin{figure}[htbp]
\centering 
\includegraphics[width=.9\textwidth]{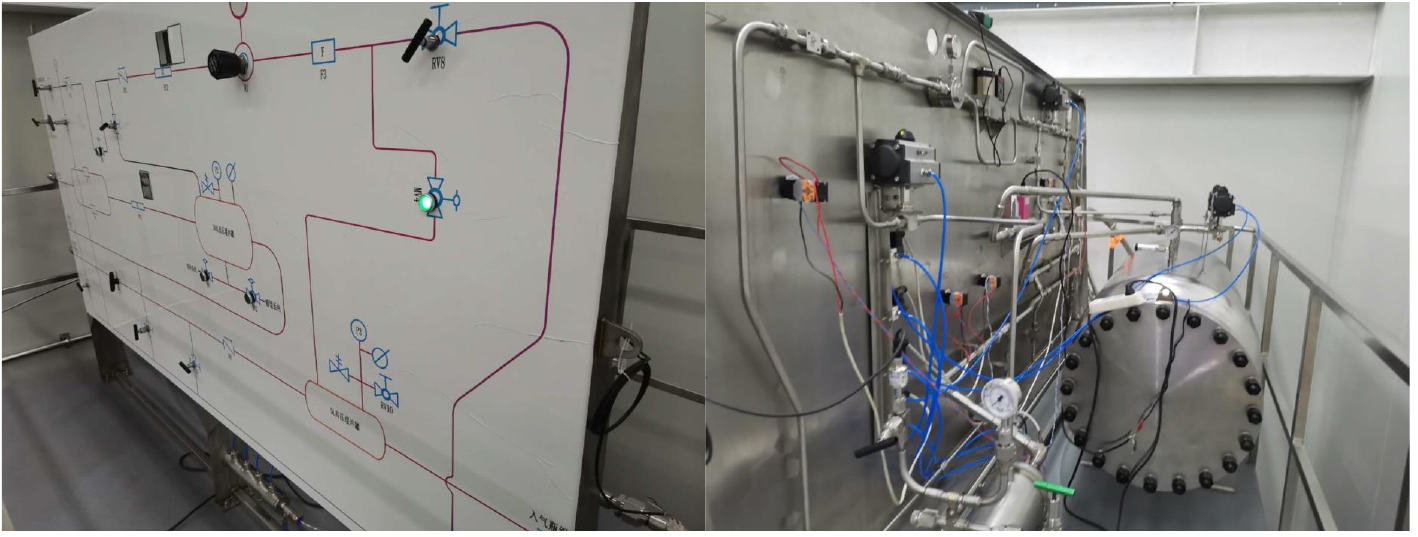}
\caption{\label{fig:2:3:2} The picture of the UHP xenon recuperation system.}
\end{figure}

The picture of PandaX-4T high speed UHP xenon recuperation system is shown in ~\ref{fig:2:3:2}.

\subsection{High-pressure bottle system}
\label{sec:2:4}

The high-pressure bottle system consists of 128 standard bottles of 40L for each. This bottle system can be filled with 6 tons of xenon at the pressure of 6 MPa at the temperature of 17 $^{\circ}$C (The density of xenon is 1464.7 $kg/m^{3}$ ).

The temperature of the xenon gas reaches to around 60 $^{\circ}$C after being compressed by the diaphragm compressor, and the dry ice is laid along the pipeline from the diaphragm compressor to the high-pressure bottle system in order to cool the temperature of the xenon gas down to 17 $^{\circ}$C. Furthermore, the high-pressure bottles are cooled down by an air conditioner. Temperature sensors (PT100) are set on the xenon tube line and the bottles to monitor the temperature.

\section{Operation process of the ultra-high purity recuperation system}
\label{sec:3}

As shown in figure ~\ref{fig:2:3:1}, xenon in the detector is vaporized by the ethanol heating circulation system at the rate of $\sim$200 SLPM, the xenon gas flows through the 0.5$\mu$m filters F1 and F2 with the pressure reduction of $ 1\times10^{5}$ Pa because of flow resistance, then enters to a buffer of 500 L to stabilize the pressure in front of the diaphragm compressor. Then the xenon is pressurized to 6 MPa, and enters a high-pressure buffer of 5 L via a check valve DV1 to avoid back flow. Two pressure sensors are settled at the buffers before and after the diaphragm compressor, respectively. Finally, the high-pressure xenon gas is filled into the high-pressure bottle system after cooling.

\begin{figure}[htbp]
\centering 
\includegraphics[width=.9\textwidth]{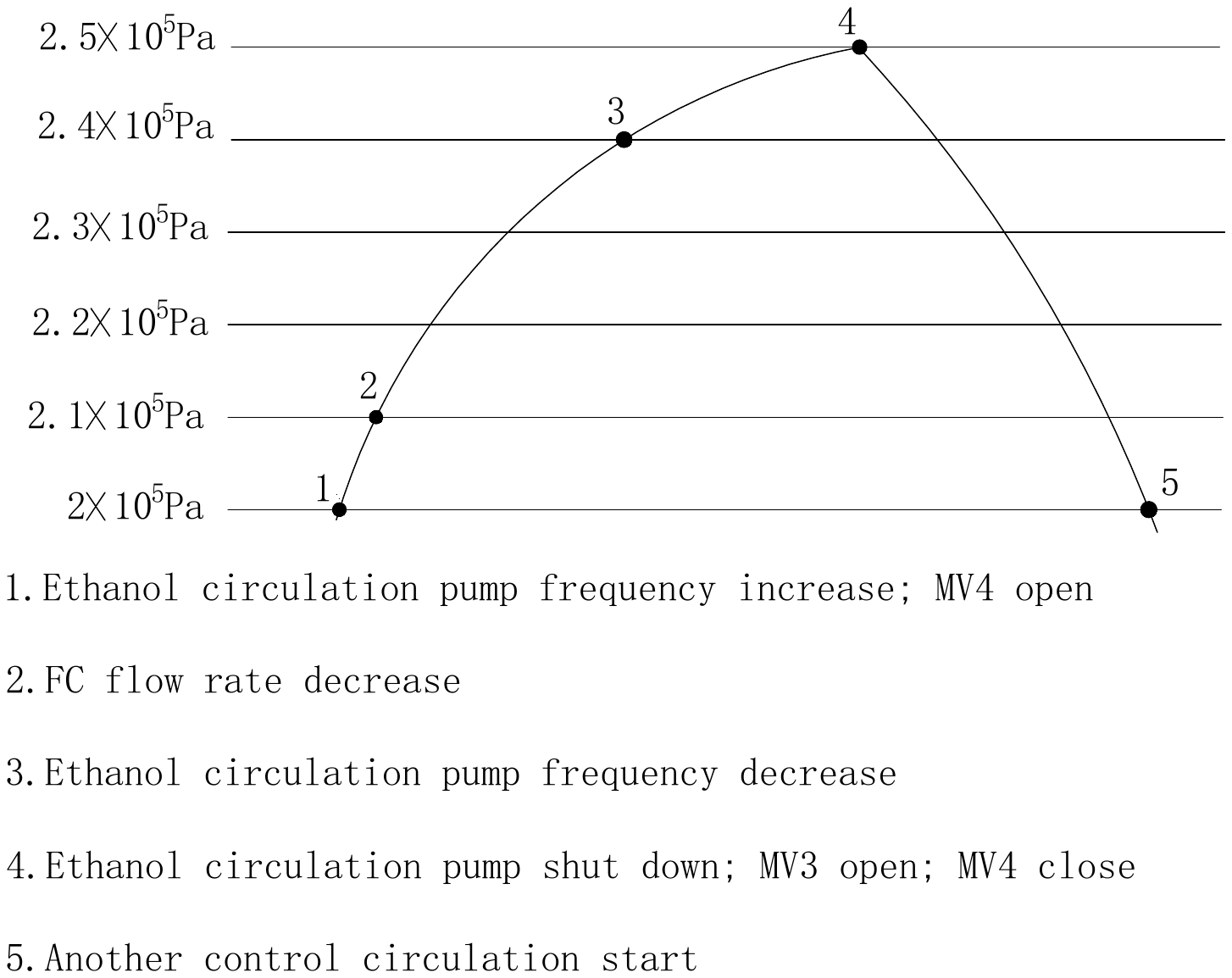}
\caption{\label{fig:3:1} The control logic of the UHP xenon recuperation system.}
\end{figure}

The operation of the Xe recuperation system is controlled by Programmable Logic Controller (PLC) mainly according to the inner pressure of the dark matter detector. The control logic is illustrated in figure ~\ref{fig:3:1}.

When the pressure in the detector is lower than $ 2\times10^{5}$ Pa, which means the recuperation rate is faster than the evaporation rate, the frequency of the ethanol circulation pump is increased by the automatic control system to rise the evaporation rate of the liquid xenon in the detector. At the same time, the pneumatic valve of the reflux pipeline MV4 is switched on automatically, and the high-pressure xenon after the diaphragm compressor flows back to the 500L buffer via a regulator to reduce the pressure of the xenon to $2\times10^{5}$ Pa and a flow controller FC (Alicat, control range: 0 $\sim$ 250 SLPM) to control the reflux rate, which is related to the pressure in the detector: the flow rate increases when the pressure in the detector decreases and vice versa.

The pressure in the detector would increase as the result of the cooperation of increasing the evaporation of the liquid xenon and rising the reflux rate of the xenon gas. The reflux rate starts to decrease when the pressure in the detector reaches to $2.1\times10^{5}$ Pa, controlled by FC through the automatic control system. The circulation rate of the ethanol heating circulation system begins to decrease when the pressure in the detector increases to $2.4\times10^{5}$ Pa.

The ethanol heating is stopped when the pressure in the detector exceeds to $2.5\times10^{5}$ Pa. At the same time, the pneumatic valve of the auxiliary recuperation system MV3 switches on automatically. Thus besides the recuperation flow rate of 200 SLPM using the diaphragm compressor, the additional xenon is recuperated by the auxiliary pipeline to the low temperature recovery bottle system at the maximum flow rate of 50 SLPM. The pressure in the detector decreases as the result of the combined operation of decreasing the evaporation of the liquid xenon and increasing the auxiliary recuperation flow rate of the xenon gas. 

By adjusting the flow rate of the ethanol heating circulation system, the reflux pipeline and the auxiliary recuperation pipeline using the automatic control system as illustrated before, the xenon gas can be compressed from 2 $\sim$ 3$\times10^{5}$ Pa to 6 MPa stably and filled into the high-pressure bottle system continuously.

\section{Operation results}
\label{sec:4}
\subsection{Xenon filling process}
\label{sec:4:1}

The pipelines of the recuperation system are vacuum pumped initially. When filling, the xenon in the dark matter detector is filled into the recuperation system to about $2\times10^{5}$ Pa before starting the diaphragm compressor. 

\begin{figure}[htbp]
\centering 
\includegraphics[width=.7\textwidth]{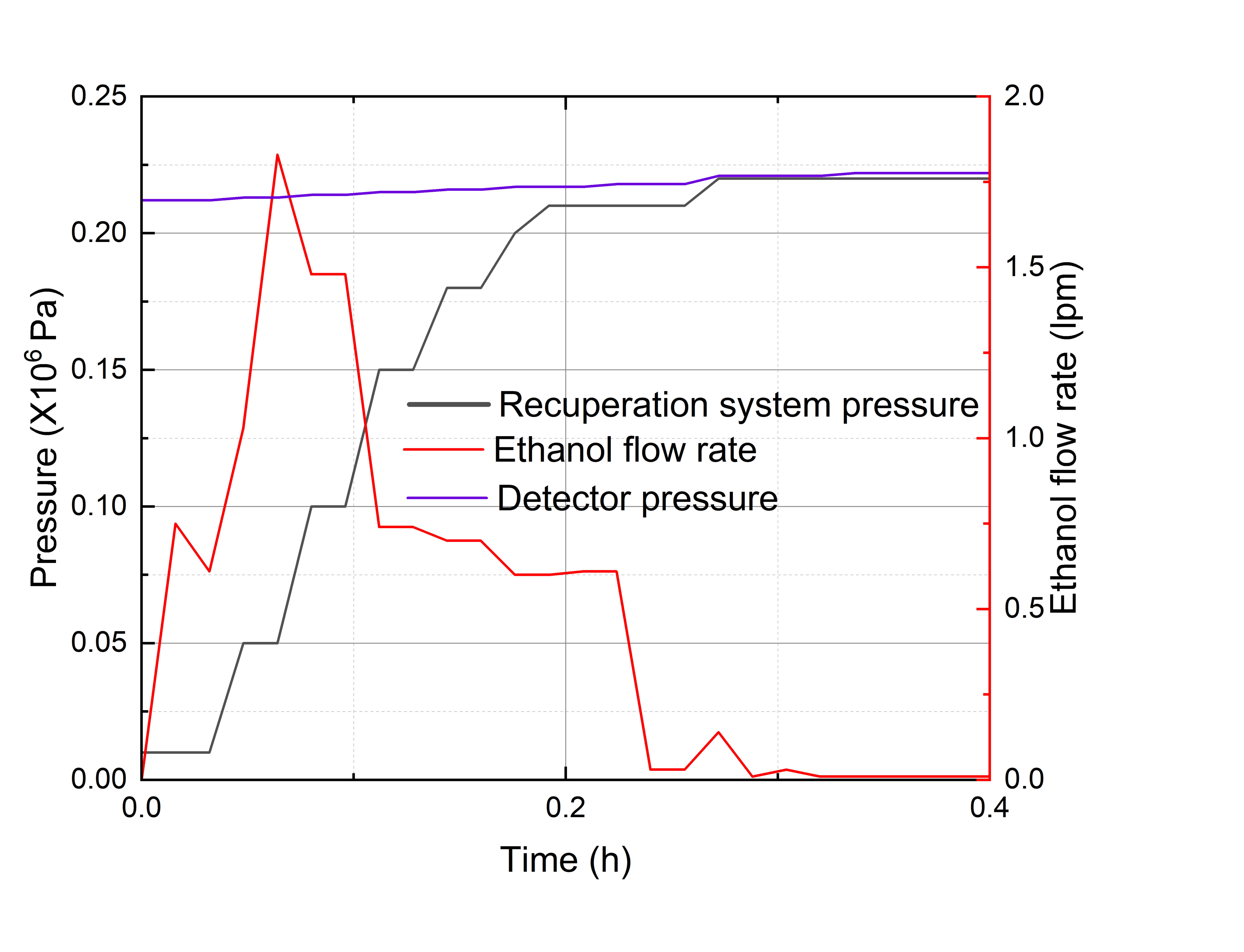}
\caption{ \label{fig:4:1}The pressure of the recuperation system and the detector, as well as the ethanol circulation rate at xenon filling process. }
\end{figure}

The pressure in the dark matter detector and the recuperation system, as well as the circulation rate of the ethanol heating system are shown in figure ~\ref{fig:4:1}. At the initial step of the ethanol heating, the temperature of the stainless steel wall of the inner vessel of the detector is 178K, so faster ethanol circulation rate is needed to heat up the stainless steel wall first, then to evaporate the liquid xenon. The maximum ethanol circulation rate of 1.83 LPM is reached at the first 5 min, then the rate drops back and stabilizes at 0.6 LPM afterwards. The pressure in the detector increases slightly from $2.1\times10^{5}$ Pa to $2.22\times10^{5}$ Pa, and the pressure in the recuperation system increases from 0 Pa (vacuum) to $2.2\times10^{5}$ Pa when the vaporized xenon gas filled into it. According to the experimental data, The recuperation system including pipelines, low-pressure buffer of 500 L, high-pressure buffer of 5 L and the high-pressure bottle system of 5.12 $m^{3}$ is filled with xenon from the detector to $2.2\times10^{5}$ Pa in 0.5 h, corresponding to 61.3 kg xenon gas in total, which means the evaporation rate is 350.8 SLPM at xenon filling process.

\subsection{Stable recuperation process}
\label{sec:4:2}

The diaphragm compressor starts when the pressure of the recuperation system reaches to 2 $\times10^{5}$ Pa. During this process, the xenon in the dark matter detector of 2 $\sim$ 3$\times10^{5}$ Pa is compressed to 6 MPa and filled into the high-pressure bottle system at the recuperation rate of 200 SLPM, with the cooperation of the control of the ethanol circulation system, the reflux pipeline and the auxiliary recuperation pipeline, following the control logic illustrated in Section 3. The gaseous space in the detector increases along the recuperation process, which makes the system safer and more stable.

\begin{figure}[htbp]
\centering 
\includegraphics[width=.7\textwidth]{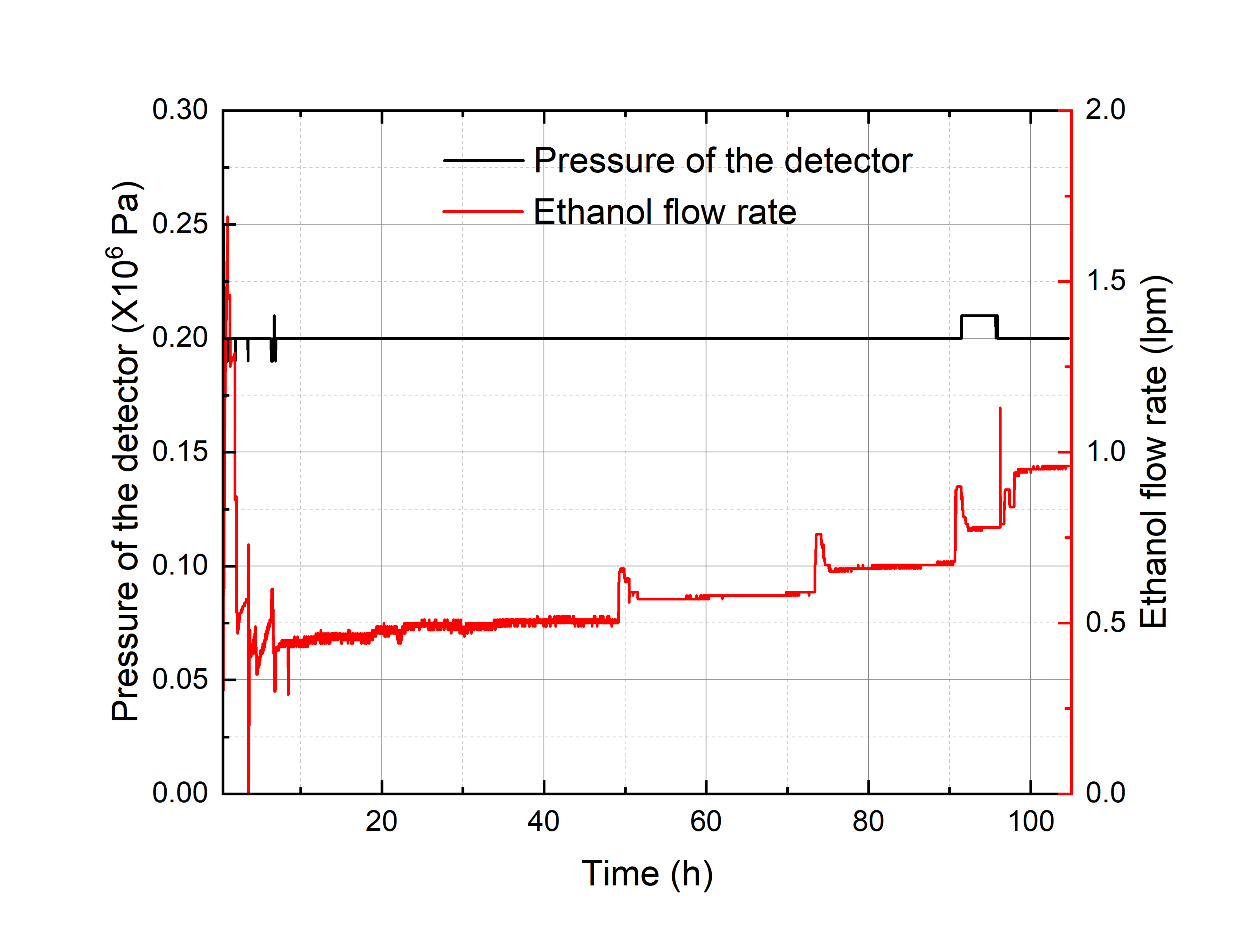}
\caption{ \label{fig:4:2} The pressure of the dark matter detector and the ethanol circulation rate at stable recuperation process. }
\end{figure}

The pressure of the detector is stable at about $2\times10^{5}$ Pa with slightly fluctuation from $1.9\times10^{5}$ Pa to $2.1\times10^{5}$ Pa under the operation of the automatic control system, as shown in figure ~\ref{fig:4:2}. The rate of the ethanol circulation system increases step by step to ensure the stable evaporation because of the decreasing of the contact area between the ethanol and the liquid xenon along the liquid xenon level decreasing.

\begin{figure}[htbp]
\centering 
\includegraphics[width=.7\textwidth]{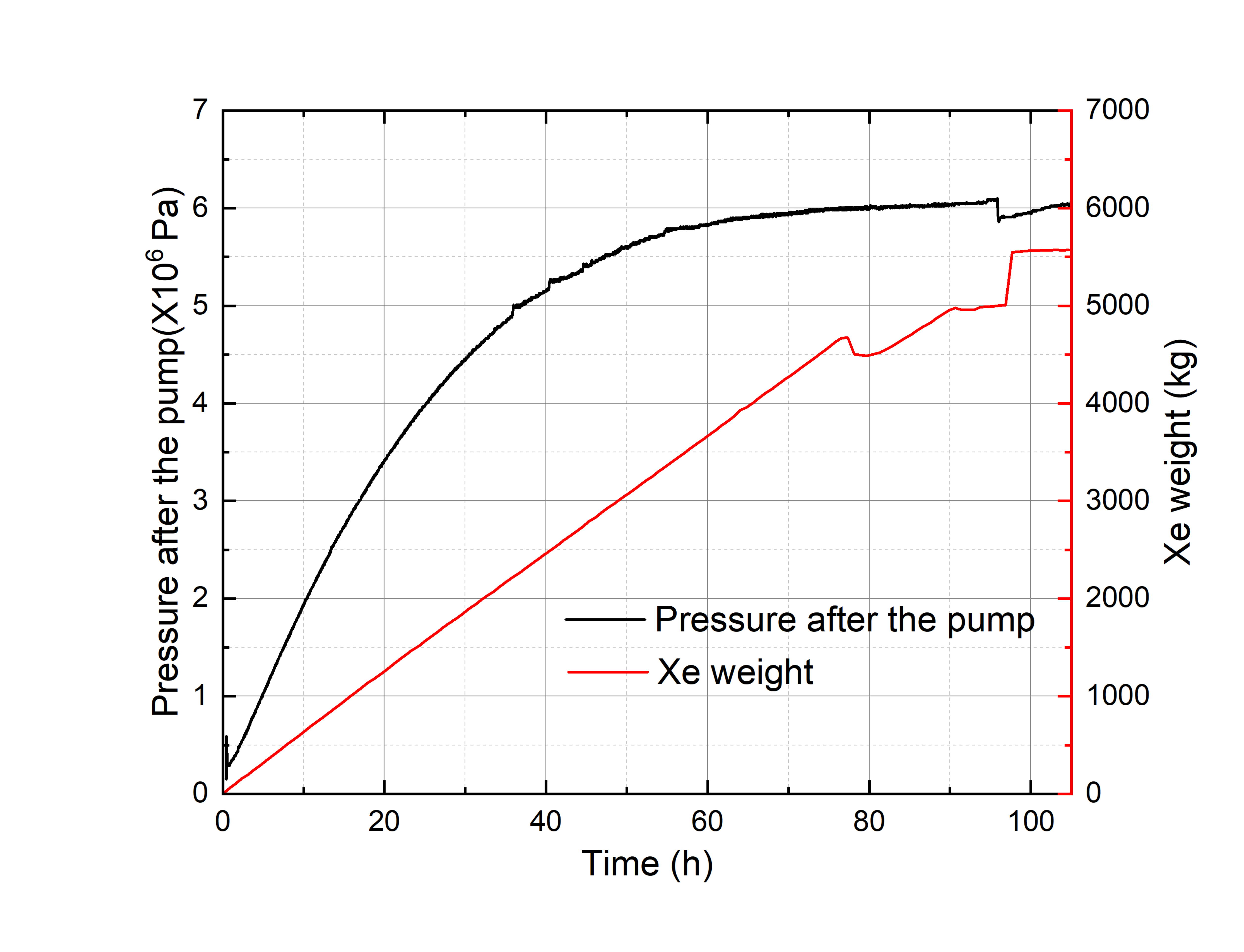}
\caption{ \label{fig:4:3} The increasing condition of the pressure and weight of the bottles at stable recuperation process. }
\end{figure}

According to figures ~\ref{fig:4:2} and ~\ref{fig:4:3}, the recuperated xenon mass and the corresponding ethanol circulation rate and liquid xenon level in the detector is listed in Table ~\ref{tab:table1}.
 
The pressure of the high-pressure bottle system is increased from 0 to 6.1 MPa stably in stable recuperation process as illustrated in figure ~\ref{fig:4:3}.

\begin{table}
\caption{\label{tab:table1}Xenon recuperation situation in stable recuperation process}
\footnotesize
\centering
\begin{tabular}{@{}llll} 
\toprule[1pt]
Time duration (h)&Ethanol circulation rate (LPM) &Recuperated xenon mass in total (t)& Liquid xenon level (m)\\
\hline
0 to 49& 0.5 & 2.4 & 2 to 1.2 \\
49 to 73& 0.6 & 3.4 & 1.2 to 0.83 \\
73 to 90& 0.7 & 4.25 & 0.83 to 0.58 \\
90 to 95& 0.8 & 4.5 & 0.58 to 0.5 \\
95 to 105& 0.9 & 5 & 0.5 to 0.33 \\
105 to 109& 1.25 & 5.75 & 0.33 to 0 \\
\bottomrule[1pt]
\end{tabular}\\
\end{table}
\normalsize

\subsection{Shut down process}
\label{sec:4:3}

 When almost all the liquid xenon is recuperated, the ethanol circulation rate increases rapidly, but the evaporation rate is not sufficient to the large recuperation flow rate of 200 SLPM. As a result, the pressure in the detector decreases continuously. It comes into shut down process when the pressure of the detector drops to $2\times10^{5}$ Pa.
 
 The ethanol circulation is switched off manually, and the valves before and after the diaphragm compressor are closed, after few minutes of the self-circulation of the diaphragm compressor, the operation of the diaphragm compressor is switched off. In the meanwhile, the valve of the auxiliary recuperation system is switched on manually, and the residual xenon in the detector is recuperated in the bottles immersed in the liquid nitrogen at the flow rate of 50 SLPM. Finally, refill the high-pressure xenon in the 5 L buffer back to the 200 L buffer via the regulator of the reflux pipeline, in order to balance the pressure before and after the compressor to avoid damaging the diaphragm.
 
\begin{figure}[htbp]
\centering 
\includegraphics[width=.7\textwidth]{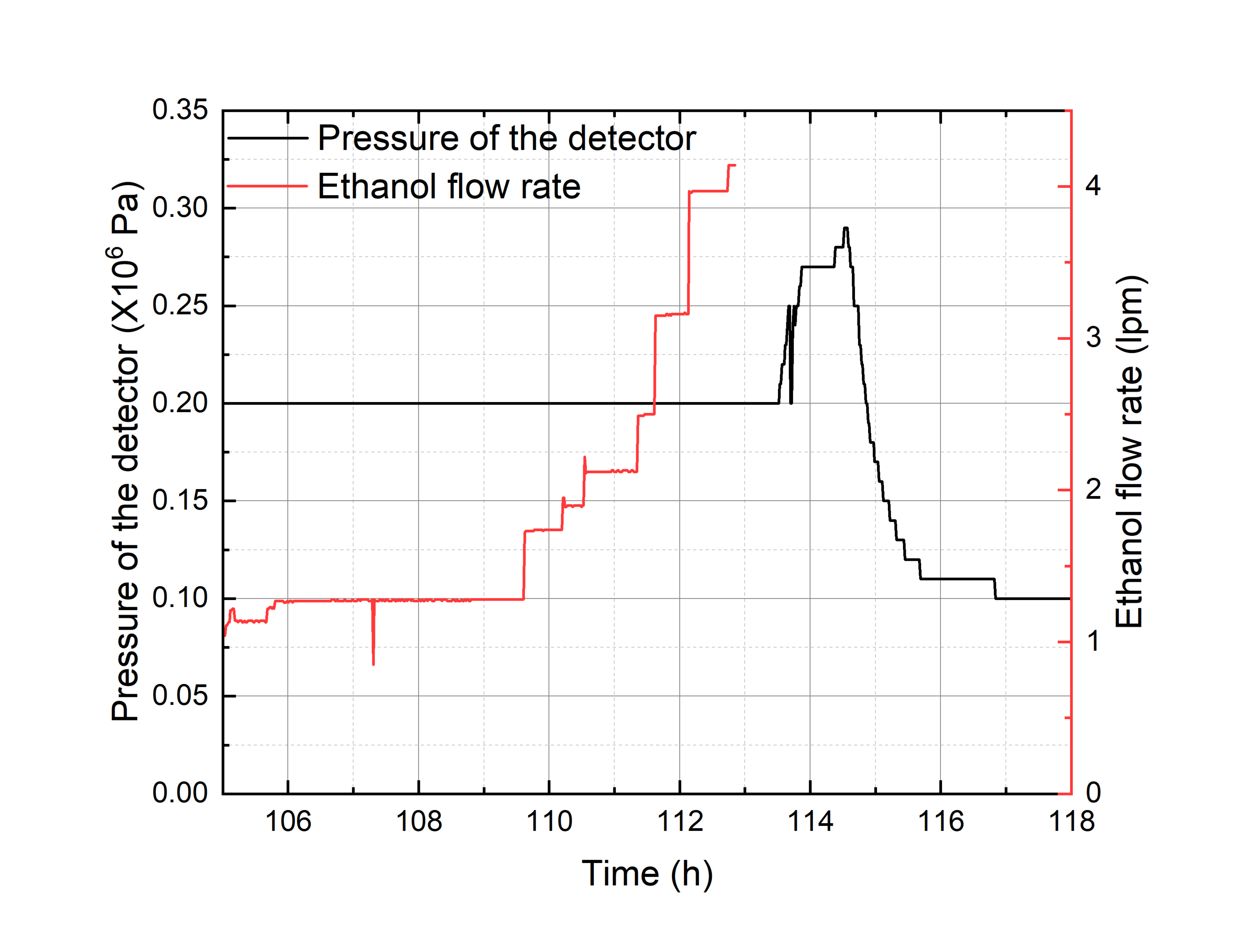}
\caption{\label{fig:4:4} The pressure of the dark matter detector and the ethanol circulation rate at shut down process.}
\end{figure}

 The pressure of the dark matter detector and the ethanol circulation rate at shut down process are shown in figure ~\ref{fig:4:4}. From the figure, the pressure of the detector increases to $2.7\times10^{5}$ Pa at the beginning of this process because of the shut down of the diaphragm compressor. Then the pressure of the detector starts to decrease after 1 h because the continuously operation of the auxiliary recuperation system. The recuperation of the xenon is completed when the pressure of the dark matter detector decreases to  $1\times10^{5}$ Pa. The Xe left inside the detector is used to keep the positive pressure, which could protect the detector from being contaminated completely if an accidental leakage happen.
 
  According to the experimental data, about 5.57 tons of Xe is recuperated by the diaphragm compressor, and the rest Xe of 26 kg is recuperated by the assistant recuperation system. 

\subsection{Purity measurement of the recuperated xenon}
\label{nt recupsec:4:4}

The recuperated xenon is stored in the high-pressure bottle system, where the xenon is sampled and measured to check if it is contaminated .

A krypton measurement system is developed and constructed for detecting krypton impurity. The krypton measurement and analysis system are based on a residual gas analyzer (RGA) and a liquid nitrogen cold trap. One of the key factors to measure the krypton concentration as low as the ppt level is the relative enrichment of krypton in xenon, which is achieved by the liquid nitrogen cold trap. The Kr concentration in the sample xenon can be obtained by analyzing the partial pressure of different component species measured by the RGA set behind the liquid nitrogen cold trap~\cite{15,16}. This krypton measurement system is calibrated by xenon samples with the known krypton concentration, which allows detecting krypton impurity at the $10^{-12}$ (ppt) level in principle. The measurement and calibration methods are described in Ref.~[17]. The gas partial pressure ratio of the $^{84}$Kr to $^{132}$Xe is a critical parameter, which reflects the krypton content. The partial pressure of $^{87}$Kr is used as background signal called "Fake" for it is a non-stable isotope, and this signal is subtracted during the analysis. The krypton concentration can be calculated by comparing the parameter $^{84}$Kr/$^{132}$Xe between the recuperated xenon and the xenon sample. Kr concentration is an important indicator to judge if there is contamination in the recuperated xenon, in which Kr should be at $10^{-12}$ level. There would be a great increase of Kr concentration if the UHP xenon was contaminated by air, in which Kr is at $10^{-7}$ level. 

\begin{figure}[htbp]
\centering 
\includegraphics[width=.9\textwidth]{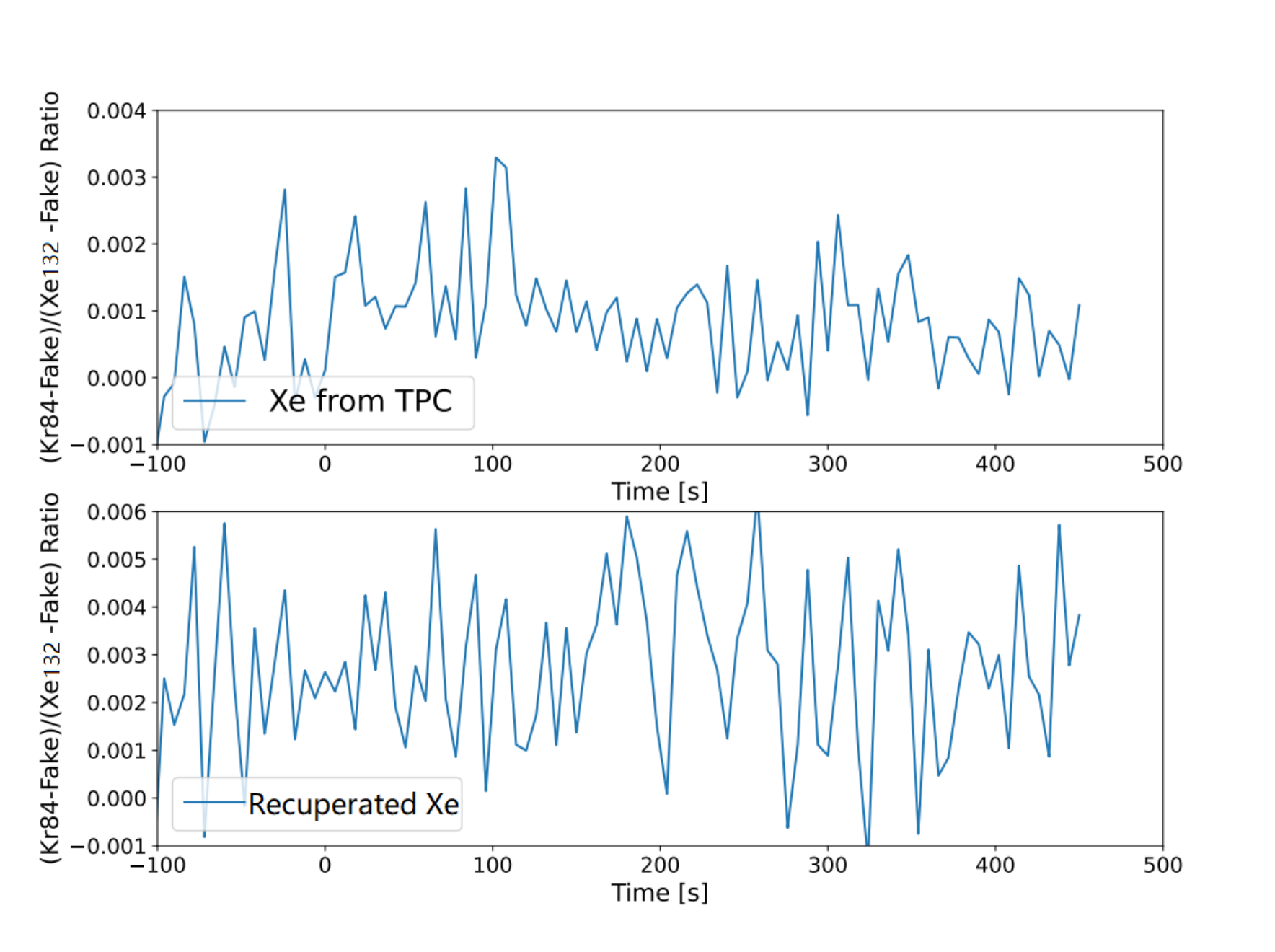}
\caption{ \label{fig:4:8} Time evolution of the partial pressure ratio of $^{84}$Kr and $^{132}$Xe of the xenon in the dark matter detector and the recuperated xenon. }
\end{figure}

The time evolution of the partial pressure ratio of $^{84}$Kr and $^{132}$Xe of the xenon in the dark matter detector and the recuperated xenon are shown in figure ~\ref{fig:4:8}. The analysis result of the Kr concentration of the xenon in the dark matter detector is less than 18.6 ppt, while the Kr concentration in the recuperated xenon sample is less than 37.6 ppt (The burn-in of the filament of the RGA causes the sensitivity deterioration of the Kr measurement system after several year’s working). From the results, it can be seen that the recuperated xenon is not contaminated.

\section{Conclusion}
\label{sec:5}

A high-speed UHP xenon recuperation system is developed to recuperate 5.6 tons of xenon from the dark matter to the high-pressure bottle system. The designed parameters and construction is described in this paper firstly. The liquid xenon in the detector is vaporized by the ethanol heating circulation system, and the gaseous xenon is compressed from 2 $\sim$ 3$\times10^{5}$ Pa (corresponding temperature of 178 K$\sim$186.5 K) to 6 MPa at the flow rate of 200 SLPM by a diaphragm compressor, then filled into high-pressure bottle system without leakage and contamination. Secondly, the operation processes of the UHP xenon recuperation system are described. The operation processes include xenon filling process, stable recuperation process and shut down process, and the duration is about 6 days, in which the stable recuperation process is 5 days. Thermodynamic analysis shows that the recuperation system performs steadily during the stable recuperation process, the recuperation system completes the technical target reliably. Finally, the recuperated xenon is sampled and measured, and the results are compared with the sampled xenon taken from the dark matter detector before the recuperation. The results show that there is no contamination during the recuperation. This system is important to the stable operation of the high-sensitivity and low background large-scale dark matter detector.

\acknowledgments

The authors would like to thank the supports of the PandaX-4T collaboration. 
This project is supported by grants from the Ministry of Science and Technology of China (No. 2016YFA0400301 and 2016YFA0400302), a Double Top-class grant from Shanghai Jiao Tong University, grants from National Science Foundation of China (Nos. 11435008, 11505112, 11525522, 11775142 and 11755001), grants from the Office of Science and Technology, Shanghai Municipal Government (Nos. 11DZ2260700, 16DZ2260200, and 18JC1410200), and the support from the Key Laboratory for Particle Physics, Astrophysics and Cosmology, Ministry of Education. 
We also thank the sponsorship from the Chinese Academy of Sciences Center for Excellence in Particle Physics (CCEPP), Hongwen Foundation in Hong Kong, and Tencent Foundation in China. Finally, we thank the CJPL administration and the Yalong River Hydropower Development Company Ltd. for indispensable logistical support and other help.


\end{document}